\documentclass[aps,pra,preprint,groupedaddress,showpacs,showkeys,floatfix]{revtex4}

\usepackage[dvips]{graphicx}
\newcommand{\bqn}{\begin{equation}}
\newcommand{\eqn}{\end{equation}}
\newcommand{\bqna}{\begin{eqnarray}}
\newcommand{\eqna}{\end{eqnarray}}
\newcommand{\bary}{\begin{array}{clcr}}
\newcommand{\eary}{\end{array}}

\begin{document}

\title
{Multiphoton ionization of the Calcium atom by linearly and
circularly polarized laser fields
}
\author{Gabriela Buica}
\affiliation
{Institute for Space Sciences, P.O. Box MG-23, Ro 77125,
Bucharest-M\u{a}gurele, Romania}

\author{Takashi Nakajima}
\email[]{t-nakajima@iae.kyoto-u.ac.jp}
\affiliation
{Institute of Advanced Energy, Kyoto University,
Gokasho, Uji, Kyoto 611-0011, Japan}

\begin{abstract}

We theoretically study multiphoton ionization of the Ca atom irradiated
by the second (photon energy 3.1 eV) and third (photon energy 4.65 eV)
harmonics of Ti:sapphire laser pulses (photon energy 1.55 eV).
Because of the dense energy level structure the second and third harmonics
of a Ti:sapphire laser are nearly single-photon
resonant with the ${4s4p}$ $^1P^o$ and ${4s5p}$ $^1P^o$ states, respectively.
Although two-photon ionization takes place through the near-resonant
intermediate states with the same symmetry in both cases, it turns out that
there are significant differences between them.
The photoelectron energy spectra exhibit the absence/presence of
substructures.
More interestingly, the photoelectron angular distributions clearly show
that the main contribution to the ionization processes by the third harmonic
arises from the far off-resonant ${4s4p}$ $^1P^o$ state rather than the
near-resonant ${4s5p}$ $^1P^o$ state. These findings can be attributed
to the fact that the dipole moment for the ${4s^2}$ $^1S^e$ - ${4s5p}$ $^1P^o$
transition is much smaller than that for the
${4s^2}$ $^1S^e$ - ${4s4p}$ $^1P^o$ transition.

\end{abstract}

\maketitle

\section{Introduction}

Above-threshold ionization (ATI) \cite{agostini} is a process
in which atoms absorb more than the minimum number of photons
required to ionize and the photoelectron energy spectrum (PES)
consists of a series of peaks that are equally separated by
the photon energy. Now the ATI and multiphoton ionization (MPI) processes
have been well-studied \cite{burnet,mauro}, especially for rare gas
atoms.
Recently, we have presented several new interesting features in the PES
of a light alkaline-earth-metal atom, Mg, interacting with a short
laser pulse \cite{gabitaka1,gabitaka3}.
We have had a close look at \textit{the intermediate ATI peaks} appearing
in the PES and clarified their origin as the \textit{off-resonant}
excitation of the bound states.
In this paper we extend our previous works on Mg to another
alkaline-earth-metal atom which is less investigated in the literature,
the Ca atom.

During the last 30 years many theoretical and experimental investigations have
been performed for Ca to obtain the atomic data and to understand its
interaction with laser fields for the {\it single-photon processes}.
The first extensive theoretical studies for the singlet series $^1S$ and $^1P$
of Ca were performed by Fischer and Hansen \cite{fischer1} with a
multi-configuration Hartree-Fock (MCFH) method which included correlations
between the valence electrons.
Later on, Mitroy  \cite{mitroy1993} calculated the energy level
and oscillator strength (OS) for the low-lying levels of Ca using the
frozen-core Hartree-Fock (FCHF) approach including a model potential.
Aymar and co-workers \cite{aymar1} extensively studied  light
alkaline-earth-metal atoms using the multichannel quantum defect theory and
eigenchannel R-matrix combined with polarization potentials.
Hansen and co-workers \cite{hansen1999} used a configuration interaction (CI)
approach, based on the B-spline basis functions and a model potential with
dielectronic core polarization potential, to calculate term energies and
wave functions for the singlet and triplet states of Ca.
Recently, Fischer and Tachiev \cite{fischer} calculated the energy levels,
transition probabilities, and lifetimes for the singlet and triplet spectra
of Ca using the MCFH method with lowest-order relativistic effects included
through a Breit-Pauli Hamiltonian.

As for the {\it MPI processes} Benec'h and Bachau \cite{bene} calculated
the one-, two-, and three-photon ionization cross sections of Ca
from the ground state with B-spline basis functions and CI procedure,
where the FCHF approach and polarization potentials were employed to construct
the atomic basis.
Regarding the experimental MPI processes of Ca, there are only several
works in the literature, all of which involve only ns and ps laser pulses.
DiMauro and co-workers \cite{kim} investigated single and double ionization
of Ca by 10 ns Nd:YAG laser pulses
at 532 and 1064 nm in the intensity range of $10^{10}- 10^{13}$W/cm$^2$.
Shao and co-workers \cite{shao} analyzed single and double ionization of Ca by
35 and 200 ps Nd:YAG laser pulses at 532 and 1064 nm in the
intensity range of $10^{10}- 2\times10^{13}$ W/cm$^2$.
Lately, Cohen and co-workers \cite{cohen} experimentally and theoretically
studied the two-photon ionization spectra of Ca in the 374-323 nm wavelength
range with a $\sim$5 ns dye laser.
Although theoretical data for MPI of alkaline-earth-metal atoms
such as Ca by fs laser pulses would provide additional valuable information
for the purpose of understanding the multiphoton ionization dynamics,
such data are still missing in the literature.

The aim of this work is to extend our previous investigations for MPI
of Mg \cite{gabitaka1,gabitaka2,gabitaka3,gabilamb} to the Ca atom:
In this paper we study the MPI processes of Ca by the second
(photon energy 3.1 eV) and third (photon energy 4.65 eV) harmonics
of Ti:sapphire laser pulses.
For this purpose we use a nonperturbative method to solve the time-dependent
Schr\"{o}dinger equation (TDSE) with two active electrons.
In Sec. \ref{I} we construct an atomic basis set in terms of discretized
states and use it in Sec. \ref{II} to numerically solve the TDSE.
Numerical results such as the ionization yields, PES, and
photoelectron angular distribution (PAD) are presented in in Sec. \ref{III}.
Atomic units (a.u.) are used throughout this paper unless otherwise mentioned.

\section{Atomic Basis States}
\label{I}

The Ca atom is a two-valence-electron atom with a closed ionic core Ca$^{2+}$
(the nucleus and the 18 inner-shell electrons $1 s^2 2s^2 2p^6 3s^2 3p^6 $)
and the two valence electrons, $4s^2$ $^1S^e$.
Since it is a heavier alkaline-earth-metal element than Be and Mg and hence
the core, Ca$^{2+}$, is softer, there are more complexities as well as
subtleties in the Ca atomic structure.
As it is already mentioned in the literature \cite{lamb0,bachau}
there are several different approaches to solve the Schr\"odinger equation
to describe the interaction of a one- and two-valence-electron atom
with a laser field.
Since the general computational procedure has already been presented in
Refs. \cite{chang,tang,chang2,gabitaka2,gabitaka3} and the specific details
about the atomic structure calculation of Ca have been reported in
recent works \cite{hansen1999,bene,mitroy}, we only make a brief description
of the method we employ.
The field-free one-electron Hamiltonian of Ca$^+$, $ H_{a}(r)$, is
expressed as:
\begin{equation}
 H_{a}(r)= -\frac{1}{2}\frac{{\rm d}^2}{{\rm d}r^2}-
            \frac{Z}{r}+\frac{l(l+1)}{2 r^2} + V_{eff}(r),
 \label{h_1e}
\end{equation}
where $ V_{eff}(r) $ is the effective potential acting on the valence
electron of Ca$^+$, ${r}$ represents the position vector of the valence
electron, $Z$ is the electric charge of the core ($Z=2$ for a
two-valence-electron atom), and $l$ is the orbital quantum number.
In our approach the effective potential, $V_{eff}(r)$, consists of
the FCHF potential (FCHFP) and the additional core-polarization potential
which will be introduced in the next subsection.

\subsection{One-Electron Orbitals: Frozen-Core Hartree-Fock approach}

To describe the ionic core Ca$^{2+}$ we have introduced the effective
potential in Eq. (\ref{h_1e}), which is a sum of the FCHFP, $V_{l}^{HF}(r) $,
and the core-polarization potential, $V_l^p(r,\alpha_s,r_l) $:
\begin{equation}
 V_{eff}(r)= V_{l}^{HF}(r) + V_l^p(r,\alpha_s,r_l).
\label{v_effhf}
\end{equation}
$V_l^p(r,\alpha_s,r_l) $
 describes the interaction between the ionic core and the valence
electrons and can be written in the form of

\begin{equation}
V_l^p(r,\alpha_s,r_l)=
-{\displaystyle\frac{\alpha_s}{2r^4}
\left[
1-\mbox{exp}^{-(r/{r_l})^6}
\right]} ,
\label{v_lp}
\end{equation}

\noindent
where $\alpha_s$ is the static dipole polarizability of Ca$^{2+}$ and
$r_l $ the cutoff radii for the different orbital angular momenta,
 $l=0,1,2,...$, etc \cite{chang}.
The  values of $r_l$ have been obtained by performing the fittings of the
one-electron energies to their experimental values \cite{nist}
for the four lowest states of $s$, $p$, $d$ and $f$ series of Ca$^+$.
We have used the following set of cutoff radii,
$r_0 = 1.5457$, $r_1= 1.5857$, $r_2 = 1.8771$, and $r_{l\ge3} = 1.5530$
together with the static dipole polarizability of Ca$^{2+}$ which is
$\alpha_s = 3.16 $ \cite{meyer}.
The relatively large value of the cutoff radius for $l=2$ is due to the
fact the $3d$ orbital penetrates the ionic core much more than the other
orbitals, and therefore the core-polarization potential is more sensitive
for the $l=2$ orbital than for other orbitals.
We note that other theoretical papers
\cite{mitroy1993,aymar1,hansen1999,mitroy,bene} use
slightly different values for the static dipole polarizability and
cutoff radii.

With the aid of the mathematical properties of the B-spline polynomials
\cite{chang, bachau} to expand the one-electron orbitals, solving the
one-electron Schr\"odinger equation for the non-relativistic one-electron
Hamiltonian given by Eq. (\ref{h_1e}) is now equivalent to an
eigenvalue problem.

\subsection{Two-electron states }

The  field-free two-electron Hamiltonian,
$ H_{a}({\bf r}_{1},{\bf r}_{2})$, can be expressed as

\begin{equation}
\label{h_2e}
H_{a} ({\bf r}_{1},{\bf r}_{2})=
\sum _{i=1}^{2}  H_a(r_i) + V(\mathbf{r}_{1},\mathbf{r}_{2}),
\label{HF}
\end{equation}

\noindent
where $ H_a(r_i) $ represents the one-electron Hamiltonian for the
$i${th} electron as shown in Eq. (\ref{h_1e}), and
$ V(\mathbf{r}_{1},\mathbf{r}_{2}) $
is a two-electron interaction operator which includes the static
Coulomb interaction $ 1/|{\bf r}_1 - {\bf r}_2| $ and the effective
dielectronic interaction potential \cite{chang,moccia}.
Here ${\bf r}_1$ and ${\bf r}_2 $ are the position vectors of the two
valence electrons.
To solve the two-electron Schr\"odinger equation for the Hamiltonian
given in Eq. (\ref{HF}) the two-electron states can be constructed
within the CI approach.
Namely, we use a linear combination of the products of the two one-electron
orbitals to represent a two-electron states and diagonalize the two-electron
Hamiltonian given in Eq.(\ref{HF}).
This is so-called a CI procedure \cite{chang,tang,chang2}.

For Ca, which is heavier than Be and Mg but still relatively light
alkaline-earth-metal atom, the $LS$ coupling is known to give a fair
description \cite{mitroy1993,hansen1999,bene,aymar1} and hence
it is sufficient to label a two-electron state by the following set of
quantum numbers:
principal, orbital, and spin quantum numbers for each electron,
$n_{i}l_{i}s_{i}$ ($i=1,2$), total orbital momentum $L$, total spin
$S$, total angular momentum $J$, and its projection $M$ on the quantization
axis.
After the CI procedure the two-electron states may be most generally labeled
by the state energy and the quantum numbers $(L,S,J,M)$, and furthermore
the above state labeling can be simplified to $(L,M)$ for singlet states
$(S=0)$.
Having obtained the two-electron wave functions we can calculate the
dipole matrix elements as well as OSs for both LP and CP laser pulses.

\section{Time-dependent Schr\"{o}dinger equation}
\label{II}

By making use of the two-electron states which have been constructed
in Sec. \ref{I}, we can solve the TDSE for the two-electron atom
interacting with a laser pulse. The TDSE reads

\begin{equation}
 i \frac{\partial}{\partial t} \Psi ({\bf r}_{1},{\bf r}_{2};t) =
 	\left[ H_{a} ({\bf r}_{1},{\bf r}_{2}) + D(t) \right]
		\Psi ({\bf r}_{1},{\bf r}_{2};t),
\label{tdse}
\end{equation}

\noindent
where $ \Psi ({\bf r}_{1},{\bf r}_{2};t)$ is the two-electron wave
function for the two electrons located at ${\bf r}_{1} $ and $ {\bf r}_{2}$
at time $ t $, and $ H_{a}({\bf r}_{1},{\bf r}_{2})$
is the field-free two-electron Hamiltonian shown in Eq. (\ref{h_2e}).
The time-dependent interaction operator, $ D(t) $, between the atom and
the laser pulse is written in the velocity gauge and dipole approximation as,

\begin{equation}
D(t) = - \textbf{A}(t) \cdot ({\bf p}_{1} + {\bf p}_{2}),
\label{dipole}
\end{equation}

\noindent
where $ {\bf p}_{1} $ and $ {\bf p}_{2} $ are the momenta of the two
electrons and $ \textbf{A}(t) $ is the vector potential of the laser field
which is given by

\begin{equation}
	\textbf{A}(t)=  \textbf{A}_0 f(t)\cos (\omega t).
\label{poten}
\end{equation}

\noindent
In the above equation $\omega$ is a photon energy and
$ \textbf A_{0} = A_{0} \textbf{e}_q $ is an amplitude of the
vector potential with $ \textbf{e}_q $ being the unit polarization vector
of the laser pulse.  The unit polarization vector is expressed in
spherical coordinates, and $ q = 0 $, $1$, and $-1$ correspond to the
LP, right-circularly polarized, and left-circularly polarized
fields, respectively.
$ f(t)$ represents the temporal envelope of the laser field
which is assumed to be a cosine-squared function, i.e.,
$ f(t)=\cos ^{2}\left( {\pi t}/{2\tau }\right) $ where $ \tau $ is the
pulse duration for the full width at half maximum (FWHM) of the vector
potential  $\textbf{A}(t)$.
The temporal integration range of TDSE in Eq. (\ref{tdse}) is taken from
$ -\tau $ to $ \tau $.

In order to solve Eq. (\ref{tdse}), the time-dependent two-electron
wave function, $\Psi ({\bf r}_{1},{\bf r}_{2};t)$, is expanded
as a linear combination of two-electron states
$\Psi ({\bf r}_{1},{\bf r}_{2};E_{n})$:

\begin{equation}
\Psi ({\bf r}_{1},{\bf r}_{2};t) =
\sum _{n, L, M}C_{E_n L M}(t) \Psi ({\bf r}_{1},{\bf r}_{2};E_{n}),
\label{wf_2e_t}
\end{equation}

\noindent
where $ C_{E_n L M}(t) $ is a time-dependent expansion coefficient for a
two-electron state with an energy, $ E_n $, an angular momentum, $L$,
and its projection on the quantization axis, $M$.
Now, by substituting Eq. (\ref{wf_2e_t}) into Eq. (\ref{tdse})
we obtain a set of first-order differential equations for the time-dependent
expansion coefficients $ C_{E_n L M}(t)$ which reads

\begin{equation}
i \frac{d}{dt} C_{E_n L M}(t) = \sum_{n',L',M'}
 	\left[ E_n \delta_{n n'} \delta_{L L'} \delta_{M M'}
	     - D_{n L M n' L' M'} (t) \right]  C_{E_{n'} L' M'}(t),
\label{C-tdse}
\end{equation}

\noindent
where $ D_{n L M n' L' M'} (t) $ represents the dipole matrix element
between two singlet states defined by the quantum numbers
$(n L M)$  and $(n' L' M')$.
This means that we have neglected the spin-forbidden transitions
between the triplet and singlet states, which is a reasonably good assumption
for a light atom such as Ca.
Specifically in what follows, we assume that the Ca atom is initially
in the ground state, $4s^2$ $^1S^e$ ($M=0$), i.e.,

\begin{equation}
| C_{E_n L M}(t = -\tau) |^2 = \delta_{n 4} \delta_{L 0} \delta_{M 0}.
\end{equation}

Once we have obtained the time-dependent expansion coefficients $C_{E_n L M}$
by solving Eq. (\ref{C-tdse}), the ionization yield, $Y$,
photoelectron energy spectrum, ${dP}/{dE_e}$, and photoelectron angular
distribution, ${dP}/{d\theta}$, can be calculated at the end of the
pulse from the following relations:

\begin{equation}
Y= 1- \sum_{n, L, M (E_n < 0)} \mid C_{E_n L M}(t = +\tau )\mid^2,
\label{yield}
\end{equation}

\noindent

\begin{equation}
\frac{dP}{dE_e}(E_e) =
 \sum_{L,M}\left|C_{E_e L M}(t=+\tau)\right|^2,
\label{pes}
\end{equation}

 and
\noindent
\begin{equation}
\frac{dP}{d\theta}(E_e,\theta)
=\left| \sum_{L,M  }
(-i)^{l_2} e^{i\delta(E_e)} \sqrt{2l_2+1}\ P_{l_2}(\cos\theta)
\ C_{E_e L M}(t=+\tau)
\right|^2,
\end{equation}

\noindent
where $E_e$ represents the photoelectron energy, $ P_{l_2}$ are the Legendre
polynomials, $l_2$ is the orbital momentum of the photoelectron, and $\theta$
 is the angle between the electric field and the photoelectron momentum
vectors.
$ \delta(E_e)$ is the total phase shift which is the sum of the Coulomb and
short-range scattering phase shifts.
The total phase shift, $ \delta(E_e)$, can be extracted from the asymptotic
behavior of the photoelectron wave function
\cite{bachau,chang,chang1,burgess} at large distances $r\to \infty$:
\begin{equation}
\Psi_{kl_2} (r) \to  \sqrt{\frac{2}{\pi k}}
\sin{\left[ k r + \frac{1}{k} \ln( 2k r )
- l_2\pi/2 + \delta (E_e) \right]},
\end{equation}
where $ k  = (2E_{e})^{1/2}$ represents the momentum of the photoelectron.
Since we employ the discretized technique to describe the wave functions
in a rigid spherical box, the photoelectron wave function vanishes
at the edge of the box ($r=R$).
This means that the following relation always holds:
\begin{equation}
k R + \frac{1}{k} \ln( 2k R ) - l_2\pi/2 + \delta (E_e)
=  m \pi,  \;\;  \mbox{where} \; m \; \mbox{is an integer},
\end{equation}
which enables us to calculate the total phase shift, $ \delta(E_e)$.

\section{Numerical results and discussion}
\label{III}

In this section we present representative numerical results for multiphoton
ionization of Ca from the ground state by the second and third harmonics
of fs Ti:sapphire laser pulses.
For the numerical calculation we have found out that a spherical box of
radius $R=500$ a.u. and the total angular momentum up to $L=9$  with
1800 states for each $L$ gives a good convergence in terms of the ionization
yield and PES.
A number of $402$ \textit{B}-spline polynomials of order $9$ with a sine-like
knot grid is employed.
Note that all the numerical results reported in this paper are calculated
for the 20 fs (FWHM) cosine-squared pulse in the velocity gauge
unless otherwise stated.

Before solving the TDSE in Eq. (\ref{tdse}), however, we must perform
several checks regarding the accuracy of the atomic basis for the singlet
states of Ca.
The level structure of the singlet states of Ca is presented in
Fig. \ref{fig1}:
The first ionization threshold lies at $E_{ion}=6.11$ eV relative to the ground
state.
In Table I we compare our calculated energies with other theoretical results
\cite{mitroy1993,hansen1999} and the experimental data for the first
ionization threshold and the first few low-lying singlet states
for $L= S, P, D$, and $F$.
The experimental data are taken from the database of
National Institute of Standards and Technology (NIST) \cite{nist}
and the energies are taken with respect to the double ionization
threshold, Ca$^{2+}$.
From Table I we notice that our results by the FCHF method provide
energy values as accurate as other theoretical results for the
ionization threshold and the first few low-lying states.

The next step is to check the accuracy of the wave function in terms
of the OSs.
Table II presents comparisons of the OSs for a single-photon absorption
we have calculated with other theoretical
\cite{mitroy1993,hansen1999} and the experimental data
\cite{nist,Parkinson76,Smith88,Smith81,Hunter87} for $L= S, P, D$, and $F$
in both length and velocity gauges for the following single-photon
transitions:
$4s^2 \;^1S^e \to  4s(4-6)p \;^1P^o $ and $3d4p \;^1P^o $,
$4s4p \;^1P^o \to  4s(5-7)s \;^1S^e $ and $4p^2 \;^1S^e $,
$4s4p \;^1P^o \to  4s(4-6)d \;^1D^e $ and $4p^2 \;^1D^e $,
$4s3d \;^1D^e \to  4s(4-6)p \;^1P^o $ and $3d4p \;^1P^o $, and finally
$4s3d \;^1D^e \to  4s(4-6)f \;^1F^o $ and $3d4p \;^1F^o $.
Clearly our results on OSs are in good agreement with other theoretical
and experimental data.
Relatively large discrepancies appear in the velocity gauge for the
$4s3d \;^1D^e  \to 4s4p \;^1P^o $ and $4s3d \;^1D^e  \to 3d4p \;^1F^o $
transitions. This may be due to the insufficient accuracy of the $3d$ orbitals.
Fortunately the discrepancies appear for the transitions with very small
values of the OSs, and therefore could hardly influence the outcome
of the TDSE calculations we report in this work.

After we have checked the accuracy of the constructed atomic basis,
we can now proceed to perform the time-dependent calculations
by solving Eq. (\ref{C-tdse}) under various intensities and photon energies
for both LP and CP laser pulses.
In the following calculations we employ $1800$ two-electron states
for each angular momentum up to $L=9$ and carry out the numerical
integration of TDSE [Eq. (\ref{C-tdse})] using a Runge-Kutta method.

\subsection{Ionization by the second harmonic of a Ti:sapphire laser}

In this subsection we investigate two-photon ionization of Ca by the second
harmonic of the Ti:sapphire laser at the photon energy $\omega = 3.1$ eV,
which is schematically shown in Fig. \ref{fig2}.
As already mentioned Ca has a relatively dense level structure and for
photons in the visible range it is quite easy to be near-resonance
with some bound states.
Indeed, the detuning from the $4s4p$ $^1P^o$ state is only $0.17$ eV,
and considering the large value of the dipole matrix element for
the transition $4s^2 \;^1S^e \to 4s4p \;^1P^o $ (Table II), we expect
a significant enhancement in the ionization signal.

Figure \ref{fig3}(a) shows the ionization yield as a function of
peak intensity for LP (solid) and CP (dashed) laser pulses.
The slope of these curves is about $1.85$ up to the peak intensity of
$I= 5 \times 10^{11}$ W/cm$^2$, after which the saturation takes place.
The calculated slope is a little bit smaller than the prediction by the
lowest order perturbation theory (LOPT) which gives a slope of 2 for
two-photon ionization processes.
Figure \ref{fig3}(b) shows the ratio between the ionization yield by CP
and  LP  laser pulses, $Y_{CP}/Y_{LP}$, as a function of peak intensity.
In the low intensity regime ($I \leq 5 \times 10^{11}$ W/cm$^{2}$) the ratio
slightly decreases with peak intensity and at peak intensity of
$ 5 \times 10^{11}$ W/cm$^2$ it is about $1.25$, which is smaller than the
ratio predicted by the perturbation theory for the two-photon ionization cross
sections of one-valence-electron atoms,
$\sigma^{(2)}_{CP}/\sigma^{(2)}_{LP}=1.4$ \cite{gont,lamb,reiss,maquet}.

In Fig. \ref{fig4} we plot the PES for the LP (solid)  and CP (dashed)
laser pulses at the peak intensity of $I= 5 \times 10^{11}$ W/cm$^2$.
The ATI peaks exhibit some small structures on the left as well on the right
wings which are equidistantly separated by the photon energy.
In order to clarify the importance of the near-resonant bound state
$ 4s4p \;^1P^o $ we solved the TDSE with the $ 4s4p \;^1P^o $ state
artificially removed during the numerical integration.
In  Fig. \ref{fig5}(a) we compare the PES (solid) with that calculated
without the state $4s4p \;^1P^o $ (dashed) in the atomic basis.
As a consequence the ionization signal is almost 4 orders of magnitude lower
and almost all substructures on the left and right wings of the ATI peaks
disappeared when $ 4s4p \;^1P^o $ is removed.
In order to see the population dynamics of near-resonant $ 4s4p \;^1P^o $,
we plot the population of $4s^2 \;^1S^e $ (circles) and $ 4s4p \;^1P^o $
(squares) states in Fig. \ref{fig5}(b) as a function of time.
Rabi oscillations take place between them.

Finally, we show in Figs. \ref{fig6}(a)-\ref{fig6}(d) the PADs at the
photoelectron energies corresponding to the first four ATI peaks
(see Fig. \ref{fig4}) by the LP pulse at the peak intensity of
$I= 5 \times 10^{11}$ W/cm$^2$.
Different ATI peaks exhibit different PADs, since different partial
waves make different contributions with different total phase shifts.
This is the reason why the PADs in Figs. \ref{fig6}(a) and \ref{fig6}(c)
and also Figs. \ref{fig6}(b) and \ref{fig6}(d) resemble each other, since
the accessible continua belong to the same parity.
We now have a closer look at Fig. \ref{fig6}(a).
The PAD has a typical profile for two-photon ionization from an
initial $S$ state with one secondary maximum at $\theta =90^\circ$ and
two minima at $\theta =54^\circ$ and $126^\circ$, respectively.
Similarly the PAD shown in Fig. \ref{fig6}(b) has a typical profile for
three-photon ionization from an initial $S$ state which exhibits
two secondary maxima and three minima.
Figures \ref{fig6}(c) and \ref{fig6}(d) present the PADs at the third and
fourth ATI peaks shown in Fig. \ref{fig4}.

\subsection{Ionization by the third harmonic of a Ti:sapphire laser}

In this subsection we study two-photon ionization by the third harmonic of the
Ti:sapphire laser at the photon energy of $4.65$ eV, which is schematically
shown in Fig. \ref{fig7}.
Although the detuning from the $4s5p$ $^1P^o$ bound state is only $0.1$ eV
which is even smaller than the case of the second harmonic in the previous
subsection, we do not expect any important enhancement in the ionization
process because the dipole moment for the transition
$4s^2 \;^1S^e \to 4s5p \;^1P^o $ is much smaller than that
for the $4s^2 \;^1S^e \to 4s4p \;^1P^o $ (see Table II).
Actually the OSs for the $4s^2 \;^1S^e \to 4s5p \;^1P^o $ transition is
anomalously small. This is in contrast with the more regular decrease of OSs
for the $3s^2 \;^1S^e \to 3snp \;^1P^o $ $(n=3-6)$ transitions of
Mg \cite{gabitaka2}.

First, we present the ionization yield in Fig. \ref{fig8}(a) as a
function of peak intensity for the LP (solid) and CP (dashed) laser pulses.
The curves for the LP and CP laser pulses look almost the same
with a slope of $1.98$ at the peak intensities lower than
$I= 2 \times 10^{13}$ W/cm$^2$, which agrees very well with the LOPT
prediction. This implies that the near-resonance with
$4s5p \;^1P^o $ makes very little contribution to the ionization yield.
In Fig. \ref{fig8}(b) the ratio between the ionization yield for CP
and LP laser pulses, $Y_{CP}/Y_{LP}$, is shown as a function of
peak intensity.
We can see that the ionization yield by the LP and CP laser pulses are
almost equal below the saturation intensity.

Figure \ref{fig9} shows the PES by the LP (solid) and CP (dashed)
laser pulses at the peak intensity of $I= 10^{13}$ W/cm$^2$.
Note the large difference of the peak intensities we have employed for
Figs. \ref{fig4} and \ref{fig9}, which, however, results in the similar
amount of the ionization yields.
Interestingly, the PES exhibits more substructures around each ATI peak,
labeled as (a)-(d) in Fig. \ref{fig9}, where substructure (b) is hard to
recognize due to the overlap with the ATI peak.
The substructures are equidistantly separated by the photon energy of
$4.65$ eV and appear for both LP and CP pulses.
These substructures in the PES resemble those we have seen for Mg
at $\omega= 4.65$ eV \cite{gabitaka1}.
The fact that the substructures appear for both LP and CP pulses
implies that the certain bound states accessible in both laser polarization
could be responsible for them.
The procedure we have employed to identify the origin of the substructures is
very similar to that we have used for Mg in our previous paper
\cite{gabitaka1}.
By inspection we expect that the bound states $4snp$ $^1P^o$ ($n$=4,5, and 6)
and $3d4p$ $^1P^o$ could generate such substructures.
Since we propagate the TDSE on the atomic basis, we can easily check this
speculation by solving the TDSE by artificially removing the particular state
under suspect, and comparing the PES with the original one obtained by
the complete atomic basis.
In Figs. \ref{fig10}(a)-\ref{fig10}(d)  we show the comparisons of the PES
calculated under the same laser parameters with those for Fig. \ref{fig9}.
We show the results obtained after the removal of a particular bound state
(dashed lines), namely (a) $4s4p$ $^1P^o$, (b) $4s5p$ $^1P^o$, (c) $4s6p$
$^1P^o$, and (d) $3d4p$ $^1P^o$ upon solving the TDSE, in comparison with the
PES for the complete calculation (solid lines) with the complete atomic basis.
When the $4s4p$ $^1P^o$ state is removed [Fig. \ref{fig10}(a)],
the substructures on the left-side of each main peak are reduced or disappear,
as highlighted by the circles.
Similarly, by removing the $4s6p$ $^1P^o$ and $3d4p$ $^1P^o$ states in
Figs. \ref{fig10}(c) and \ref{fig10}(d), another small spikes labeled as (c)
and (d) in Fig. \ref{fig9} disappear.
As for Fig. \ref{fig10}(b) by artificially removing the
\textit{near-resonant} state $4s5p$ $^1P^o$, the height of the main ATI peaks
is only slightly reduced since the state $4s5p$ $^1P^o$ brings a small
contribution in the ionization process (see Table II).
These comparisons indicate that the physical origin of the substructures
labeled as (a)-(d) in Fig. \ref{fig9} is quite similar to that
we have already found for the singlet as well triplet states of Mg
\cite{gabitaka1,gabitaka3}.
Briefly, the \textit{off-resonant} bound states such as  $4s4p$ $^1P^o$,
 $4s6p$ $^1P^o$, and  $3d4p$ $^1P^o$ and the \textit{near-resonant}
bound state $4s5p$ $^1P^o$  are the origin of the substructures.

Finally, we show in Figs. \ref{fig11}(a)-\ref{fig11}(d) the PADs for the
first four ATI peaks in Fig. \ref{fig9} by the LP pulse at the peak intensity
of $I= 10^{12}$ W/cm$^2$.
The peak intensity is chosen to be low to avoid any undesired
complications at higher intensities.
Again, different ATI peaks result in the different PADs.
In order to examine the influence of the intermediate bound states
with the $^1P^o$ symmetry on the PAD we make a comparison of PADs
by artificially removing the $4s4p$ $^1P^o$ (dashed), $4s5p$ $^1P^o$
(dot-dashed), $4s6p$  $^1P^o$ (dot-dot-dashed), and $3d4p$ $^1P^o$
(dot-dotted) states. The results are shown in Fig. \ref{fig12}.
It turns out that the PAD is very sensitive to the removal of the
\textit{off-resonant} $4s4p$ $^1P^o$ state (see Fig. \ref{fig7}),
while the change is very little when other bound states, including
the near-resonant $4s5p$ $^1P^o$ state, are removed.

\section{Conclusions}

We have theoretically studied multiphoton ionization of Ca by linearly and
circularly polarized fs laser pulses at the photon energies of 3.1 eV and
4.65 eV in terms of the ionization yields, photoelectron energy spectra,
and photoelectron angular distributions.
At the photon energy of $\omega= 3.1$ eV, the ionization process is strongly
enhanced due to the presence of the near-resonant $4s4p$ $^1P^o$ state
which has a large dipole moment from the ground state, and the
ionization yield is about four orders of magnitude larger compared with
a case without a resonance.
The photoelectron energy spectrum hardly shows substructures, because
any possible substructures are buried between the strongly enhanced ATI peaks.
In contrast, at the photon energy of $\omega= 4.65$ eV, the photoelectron
energy spectrum exhibits many substructures due to the real excitation of the
\textit{near-resonant} ${4s5p}$ $^1P^o$ state and some
\textit{off-resonant} bound states such as ${4s4p}$ $^1P^o$,
${4s6p}$ $^1P^o$, and $3d4p$ $^1P^o$, ... etc.
Interestingly, the far off-resonant ${4s4p}$ $^1P^o$ state still makes a
very large contribution to the ionization processes in this case,
which can be most clearly understood in terms of the photoelectron angular
distribution.

\acknowledgments{
G.B. acknowledges hospitality from the Institute of Advanced Energy,
Kyoto University during her stay.
The work by G.B. and T.N. was respectively supported by a research program
from the LAPLAS 3 and CNCSIS contract No. 558/2009 and a Grant-in-Aid for
scientific research from the Ministry of Education and Science of Japan.
}

\newpage

\samepage{
\begin{center}
\begin{tabular}{lp{5in}}
Table I. & Comparison of the energies for the first ionization threshold
and the first few low-lying states of Ca.
The energies are expressed in eV with respect to the energy of Ca$^{2+}$.
\\
\end{tabular}
\begin{tabular}{c c c c c c c c c c c c c c c c c c c c c c c c }
\hline \hline
     \multicolumn{1}{c}{$  $}
&&&&  \multicolumn{1}{c}{ ${\rm {Present} } $}
&&&&  \multicolumn{1}{c}{ {\rm Theory \cite{mitroy1993} } }
&&&&  \multicolumn{1}{c}{ {\rm Theory \cite{hansen1999} } }
&&&&  \multicolumn{1}{c}{ {\rm NIST  \cite{nist} }  }
\\
\hline\hline
  \multicolumn{1}{c}{$ E_{{\rm Ca}^+(4s)} $}
&&&&  \multicolumn{1}{c}{$ -11.87199  $}
&&&&  \multicolumn{1}{c}{$ -11.87112  $}
&&&&  \multicolumn{1}{c}{$ -11.87179  $}
&&&&  \multicolumn{1}{c}{$ -11.87172  $}
 \\
\hline \hline
     \multicolumn{1}{c}{$ E_{4s^2 \;^1S^e}  $}
&&&&  \multicolumn{1}{c}{$ -18.011   $}
&&&&  \multicolumn{1}{c}{$ -17.950   $}
&&&&  \multicolumn{1}{c}{$ -17.988   $}
&&&&  \multicolumn{1}{c}{$ -17.98488 $}
 \\
\hline
  \multicolumn{1}{c}{$  E_{4s5s \;^1S^e }  $}
&&&&  \multicolumn{1}{c}{$ -13.859  $}
&&&&  \multicolumn{1}{c}{$ -13.842  $}
&&&&  \multicolumn{1}{c}{$ -13.857  $}
&&&&  \multicolumn{1}{c}{$ -13.85406$}
 \\
\hline
  \multicolumn{1}{c}{$ E_{4s6s \;^1S^e }  $}
&&&&  \multicolumn{1}{c}{$ -12.973  $}
&&&&  \multicolumn{1}{c}{$ -12.920  $}
&&&&  \multicolumn{1}{c}{$ -12.950  $}
&&&&  \multicolumn{1}{c}{$ -12.93991$}
\\
\hline
  \multicolumn{1}{c}{$ E_{4p^2 \;^1S^e }   $}
&&&&  \multicolumn{1}{c}{$ -12.796  $}
&&&&  \multicolumn{1}{c}{$ -12.739  $}
&&&&  \multicolumn{1}{c}{$ -12.793  $}
&&&&  \multicolumn{1}{c}{$ -12.80404$}
  \\
\hline \hline
     \multicolumn{1}{c}{$ E_{4s4p \;^1P^o}  $}
&&&&  \multicolumn{1}{c}{$ -15.065  $}
&&&&  \multicolumn{1}{c}{$ -15.063  $}
&&&&  \multicolumn{1}{c}{$ -15.071  $}
&&&&  \multicolumn{1}{c}{$ -15.05236$}
 \\
\hline
  \multicolumn{1}{c}{$ E_{4s5p \;^1P^o}$}
&&&&  \multicolumn{1}{c}{$ -13.427   $}
&&&&  \multicolumn{1}{c}{$ -13.429   $}
&&&&  \multicolumn{1}{c}{$ -13.436   $}
&&&&  \multicolumn{1}{c}{$ -13.43073 $}
 \\
\hline
  \multicolumn{1}{c}{$  E_{4s6p \;^1P^o}    $}
&&&&  \multicolumn{1}{c}{$ -12.798  $}
&&&&  \multicolumn{1}{c}{$ -12.805  $}
&&&&  \multicolumn{1}{c}{$ -12.821  $}
&&&&  \multicolumn{1}{c}{$ -12.81733$}
\\
\hline
  \multicolumn{1}{c}{$  E_{3d4p \;^1P^o}    $}
&&&&  \multicolumn{1}{c}{$ -12.512  $}
&&&&  \multicolumn{1}{c}{$ -12.522  $}
&&&&  \multicolumn{1}{c}{$ -12.543  $}
&&&&  \multicolumn{1}{c}{$ -12.53782$}
  \\
\hline \hline
     \multicolumn{1}{c}{$  E_{4s3d \;^1D^e}  $}
&&&&  \multicolumn{1}{c}{$ -15.217   $}
&&&&  \multicolumn{1}{c}{$ -15.230   $}
&&&&  \multicolumn{1}{c}{$ -15.286   $}
&&&&  \multicolumn{1}{c}{$ -15.27586 $}
 \\
\hline
  \multicolumn{1}{c}{$  E_{4s4d \;^1D^e}    $}
&&&&  \multicolumn{1}{c}{$ -13.321  $}
&&&&  \multicolumn{1}{c}{$ -13.322  $}
&&&&  \multicolumn{1}{c}{$ -13.377  $}
&&&&  \multicolumn{1}{c}{$ -13.36048$}
 \\
\hline
  \multicolumn{1}{c}{$  E_{4p^2 \;^1D^e}    $}
&&&&  \multicolumn{1}{c}{$ -12.901  $}
&&&&  \multicolumn{1}{c}{$ -12.914  $}
&&&&  \multicolumn{1}{c}{$ -12.939  $}
&&&&  \multicolumn{1}{c}{$ -12.93626$}
\\
\hline
  \multicolumn{1}{c}{$  E_{4s5d \;^1D^e}    $}
&&&&  \multicolumn{1}{c}{$ -12.658  $}
&&&&  \multicolumn{1}{c}{$ -12.650  $}
&&&&  \multicolumn{1}{c}{$ -12.666  $}
&&&&  \multicolumn{1}{c}{$ -12.66359$}
  \\
\hline \hline
     \multicolumn{1}{c}{$  E_{3d4p \;^1F^o} $}
&&&&  \multicolumn{1}{c}{$ -12.925  $}
&&&&  \multicolumn{1}{c}{$ -12.934  $}
&&&&  \multicolumn{1}{c}{$ -12.964  $}
&&&&  \multicolumn{1}{c}{$ -12.95882$}
 \\
\hline
  \multicolumn{1}{c}{$  E_{4s4f \;^1F^o}    $}
&&&&  \multicolumn{1}{c}{$ -12.728  $}
&&&&  \multicolumn{1}{c}{$ -12.723  $}
&&&&  \multicolumn{1}{c}{$ -12.737  $}
&&&&  \multicolumn{1}{c}{$ -12.73494$}
 \\
\hline
  \multicolumn{1}{c}{$  E_{4s5f \;^1F^o}    $}
&&&&  \multicolumn{1}{c}{$ -12.428  $}
&&&&  \multicolumn{1}{c}{$ -12.424  $}
&&&&  \multicolumn{1}{c}{$ -12.430  $}
&&&&  \multicolumn{1}{c}{$ -12.42978$}
\\
\hline
  \multicolumn{1}{c}{$  E_{4s6f \;^1F^o}    $}
&&&&  \multicolumn{1}{c}{$ -12.258  $}
&&&&  \multicolumn{1}{c}{$ -12.255  $}
&&&&  \multicolumn{1}{c}{$ -12.259  $}
&&&&  \multicolumn{1}{c}{$ -12.25899$}
  \\   \hline\hline\\
\end{tabular}
\end{center}

\newpage

\begin{center}
\begin{tabular}{lp{6.in}}
Table II. & Comparison of the absorption oscillator strengths
(in atomic units and length/velocity gauge) between the few representative
bound states with $^1S^e$,$^1P^o$, $^1D^e$, and $^1F^o$ symmetries.
\\
\end{tabular}
\begin{tabular}{c c c c c c c c c c c c c c }
\hline  \hline
     \multicolumn{1}{c}{$ 4s^2 \;^1S^e\rightarrow$}
&&&  \multicolumn{1}{c}{$ 4s4p \;^1P^o $}
&&&  \multicolumn{1}{c}{$ 4s5p \;^1P^o $}
&&&  \multicolumn{1}{c}{$ 4s6p \;^1P^o $}
&&&  \multicolumn{1}{c}{$ 3d4p \;^1P^o $}
 \\
\hline
     \multicolumn{1}{c}{$ {\rm {Present}} $}
&&&  \multicolumn{1}{c}{$1.921/1.756 $}
&&&  \multicolumn{1}{c}{$4.115/3.563[-3]$}
&&&  \multicolumn{1}{c}{$2.677/2.151[-2]$}
&&&  \multicolumn{1}{c}{$7.496/5.056[-2]$}
\\
\hline
     \multicolumn{1}{c}{ {\rm Theory }\cite{mitroy1993} }
&&&  \multicolumn{1}{c}{$1.82/1.781   $}
&&&  \multicolumn{1}{c}{$1.08/2.30[-3]$}
&&&  \multicolumn{1}{c}{$3.72/3.14[-2]$}
&&&  \multicolumn{1}{c}{$7.27/6.66[-2]$}
 \\
\hline
 \multicolumn{1}{c}{ {\rm Theory} \cite{hansen1999} }
&&&  \multicolumn{1}{c}{$ 1.745 $}
&&&  \multicolumn{1}{c}{$ 1.95[-3]$}
&&&  \multicolumn{1}{c}{$ 3.62[-2]$}
&&&  \multicolumn{1}{c}{$ 6.46[-2]$}
\\
\hline
    \multicolumn{1}{c}{ {\rm NIST } \cite{nist}   }
&&&  \multicolumn{1}{c}{$ 1.75$}
&&&  \multicolumn{1}{c}{$     $}
&&&  \multicolumn{1}{c}{$ 4.32 [-2] $}
&&&  \multicolumn{1}{c}{$ 7.01 [-2]$}
\\
\hline
    \multicolumn{1}{c}{ {\rm Experimental } \cite{Parkinson76} }
&&&  \multicolumn{1}{c}{$ 1.75     $}
&&&  \multicolumn{1}{c}{$ 9.0 [-4] $}
&&&  \multicolumn{1}{c}{$ 4.1 [-2] $}
&&&  \multicolumn{1}{c}{$ 6.6 [-2] $}
  \\
\hline \hline
     \multicolumn{1}{c}{$ 4s4p \;^1P^o\rightarrow$}
&&&  \multicolumn{1}{c}{$ 4s5s \;^1S^e $}
&&&  \multicolumn{1}{c}{$ 4s6s \;^1S^e $}
&&&  \multicolumn{1}{c}{$ 4p^2 \;^1S^e $}
&&&  \multicolumn{1}{c}{$ 4s7s \;^1S^e $}
\\
\hline
     \multicolumn{1}{c}{$ {\rm {Present}} $}
&&&  \multicolumn{1}{c}{$ 0.112/0.091$}
&&&  \multicolumn{1}{c}{$ 2.189/3.536[-2]$}
&&&  \multicolumn{1}{c}{$ 0.110/0.129    $}
&&&  \multicolumn{1}{c}{$ 1.648/1.655[-2]$}
\\
\hline
     \multicolumn{1}{c}{ {\rm Theory }\cite{mitroy1993} }
&&&  \multicolumn{1}{c}{$ 0.118/0.114  $}
&&&  \multicolumn{1}{c}{$ 0.909/2.26[-2]$}
&&&  \multicolumn{1}{c}{$ 0.120/0.148   $}
&&&  \multicolumn{1}{c}{$  $}
 \\
\hline
    \multicolumn{1}{c}{ {\rm Theory } \cite{hansen1999} }
&&&  \multicolumn{1}{c}{$ 0.128   $}
&&&  \multicolumn{1}{c}{$ 7.50 [-3]$}
&&&  \multicolumn{1}{c}{$ 0.116   $}
&&&  \multicolumn{1}{c}{$ 1.01 [-2]$}
\\
\hline
     \multicolumn{1}{c}{ {\rm Experimental } \cite{Smith88}}
&&&  \multicolumn{1}{c}{$  $}
&&&  \multicolumn{1}{c}{$ 9.0[-3]    $}
&&&  \multicolumn{1}{c}{$ 0.114      $}
&&&  \multicolumn{1}{c}{$ 1.33[-2]   $}
 \\
\hline \hline
     \multicolumn{1}{c}{$ 4s4p \;^1P^o\rightarrow$}
&&&  \multicolumn{1}{c}{$ 4s4d \;^1D^e $}
&&&  \multicolumn{1}{c}{$ 4p^2 \;^1D^e $}
&&&  \multicolumn{1}{c}{$ 4s5d \;^1D^e $}
&&&  \multicolumn{1}{c}{$ 4s6d \;^1D^e $}
\\
\hline
     \multicolumn{1}{c}{$ {\rm {Present}} $}
&&&  \multicolumn{1}{c}{$ 0.229/0.198 $}
&&&  \multicolumn{1}{c}{$ 0.573/0.604 $}
&&&  \multicolumn{1}{c}{$ 0.299/0.298$}
&&&  \multicolumn{1}{c}{$ 5.768/ 5.814[-2]$}
 \\
\hline
     \multicolumn{1}{c}{ {\rm Theory} \cite{mitroy1993} }
&&&  \multicolumn{1}{c}{$ 0.193/0.164$}
&&&  \multicolumn{1}{c}{$ 0.550/0.531    $}
&&&  \multicolumn{1}{c}{$ 0.283/0.293    $}
&&&  \multicolumn{1}{c}{$ 5.28/5.38 [-2] $}
 \\
\hline
     \multicolumn{1}{c}{ {\rm Theory}\cite{hansen1999} }
&&&  \multicolumn{1}{c}{$ 0.206     $}
&&&  \multicolumn{1}{c}{$ 0.548     $}
&&&  \multicolumn{1}{c}{$ 0.265     $}
&&&  \multicolumn{1}{c}{$ 4.527 [-2]$}
 \\
\hline
     \multicolumn{1}{c}{ {\rm Experimental }\cite{Smith88}}
&&&  \multicolumn{1}{c}{$ 0.207   $}
&&&  \multicolumn{1}{c}{$ 0.58    $}
&&&  \multicolumn{1}{c}{$ 0.28    $}
&&&  \multicolumn{1}{c}{$ 4.4[-2] $}
\\
\hline \hline
     \multicolumn{1}{c}{$ 4s3d \;^1D^e\rightarrow$}
&&&  \multicolumn{1}{c}{$ 4s4p \;^1P^o $}
&&&  \multicolumn{1}{c}{$ 4s5p \;^1P^o $}
&&&  \multicolumn{1}{c}{$ 4s6p \;^1P^o $}
&&&  \multicolumn{1}{c}{$ 3d4p \;^1P^o $}
\\
\hline
     \multicolumn{1}{c}{$ {\rm {Present}} $}
&&&  \multicolumn{1}{c}{$ 1.14[-3]/5.936[-2] $}
&&&  \multicolumn{1}{c}{$ 5.730/7.876 [-2]$}
&&&  \multicolumn{1}{c}{$ 6.112/6.756 [-2]$}
&&&  \multicolumn{1}{c}{$ 5.378/5.171 [-2]$}
 \\
\hline
     \multicolumn{1}{c}{ {\rm Theory} \cite{mitroy1993}  }
&&&  \multicolumn{1}{c}{$ 8.759[-4]/6.332[-2] $}
&&&  \multicolumn{1}{c}{$ 5.503/ 7.619   [-2] $}
&&&  \multicolumn{1}{c}{$ 5.533/ 6.179   [-2] $}
&&&  \multicolumn{1}{c}{$ 5.26/  5.73    [-2] $}
 \\
\hline
 \multicolumn{1}{c}{ {\rm Theory} \cite{hansen1999} }
&&&  \multicolumn{1}{c}{$ 1.46 [-3]$}
&&&  \multicolumn{1}{c}{$ 5.85 [-2]$}
&&&  \multicolumn{1}{c}{$ 6.568[-2]$}
&&&  \multicolumn{1}{c}{$ 5.166[-2]$}
 \\
\hline
     \multicolumn{1}{c}{ {\rm NIST} \cite{nist}}
&&&  \multicolumn{1}{c}{$      $}
&&&  \multicolumn{1}{c}{$  4.9 [-2] $}
&&&  \multicolumn{1}{c}{$  7.5 [-2] $}
&&&  \multicolumn{1}{c}{$  7.6 [-2] $}
 \\
\hline
     \multicolumn{1}{c}{ {\rm Experimental } \cite{Smith81,Hunter87}}
&&&  \multicolumn{1}{c}{$  1.02  [-3]   $}
&&&  \multicolumn{1}{c}{$  5.985  [-2]  $}
&&&  \multicolumn{1}{c}{$  5.66  [-2]   $}
&&&  \multicolumn{1}{c}{$  6.76  [-2]   $}
\\
\hline \hline
     \multicolumn{1}{c}{$ 4s3d \;^1D^e\rightarrow$}
&&&  \multicolumn{1}{c}{$ 3d4p \;^1F^o $}
&&&  \multicolumn{1}{c}{$ 4s4f \;^1F^o $}
&&&  \multicolumn{1}{c}{$ 4s5f \;^1F^o $}
&&&  \multicolumn{1}{c}{$ 4s6f \;^1F^o $}
\\
\hline
     \multicolumn{1}{c}{$ {\rm {Present}} $}
&&&  \multicolumn{1}{c}{$ 6.877/3.397 [-2]$}
&&&  \multicolumn{1}{c}{$ 0.140/0.117    $}
&&&  \multicolumn{1}{c}{$ 5.657/5.161 [-2]$}
&&&  \multicolumn{1}{c}{$ 3.053/2.842 [-2]$}
 \\
\hline
     \multicolumn{1}{c}{ {\rm Theory}\cite{mitroy1993}}
&&&  \multicolumn{1}{c}{$ 9.25/ 5.45[-2]$}
&&&  \multicolumn{1}{c}{$ 0.129/0.111   $}
&&&  \multicolumn{1}{c}{$ 4.99 /4.56[-2]$}
&&&  \multicolumn{1}{c}{$ 2.23 /2.27[-2]$}
 \\
\hline
     \multicolumn{1}{c}{ {\rm Theory}\cite{hansen1999} }
&&&  \multicolumn{1}{c}{$ 7.144 [-2]$}
&&&  \multicolumn{1}{c}{$ 0.131     $}
&&&  \multicolumn{1}{c}{$ 5.57  [-2]$}
&&&  \multicolumn{1}{c}{$ 3.038 [-2]$}
 \\
\hline
     \multicolumn{1}{c}{ {\rm NIST } \cite{nist} }
&&&  \multicolumn{1}{c}{$ 9.8 [-2] $}
&&&  \multicolumn{1}{c}{$ 9.39[-2] $}
&&&  \multicolumn{1}{c}{$ 7.6 [-2] $}
&&&  \multicolumn{1}{c}{$ 3.2 [-2] $}
\\
\hline
     \multicolumn{1}{c}{ {\rm Experimental }\cite{Smith81}}
&&&  \multicolumn{1}{c}{$ 9.8 [-2]  $}
&&&  \multicolumn{1}{c}{$ 0.138     $}
&&&  \multicolumn{1}{c}{$ $}
&&&  \multicolumn{1}{c}{$  $}
 \\
\hline\hline
\end{tabular}

\end{center}

\newpage
\begin{figure}

\centering
\includegraphics[width=6in,angle=0]{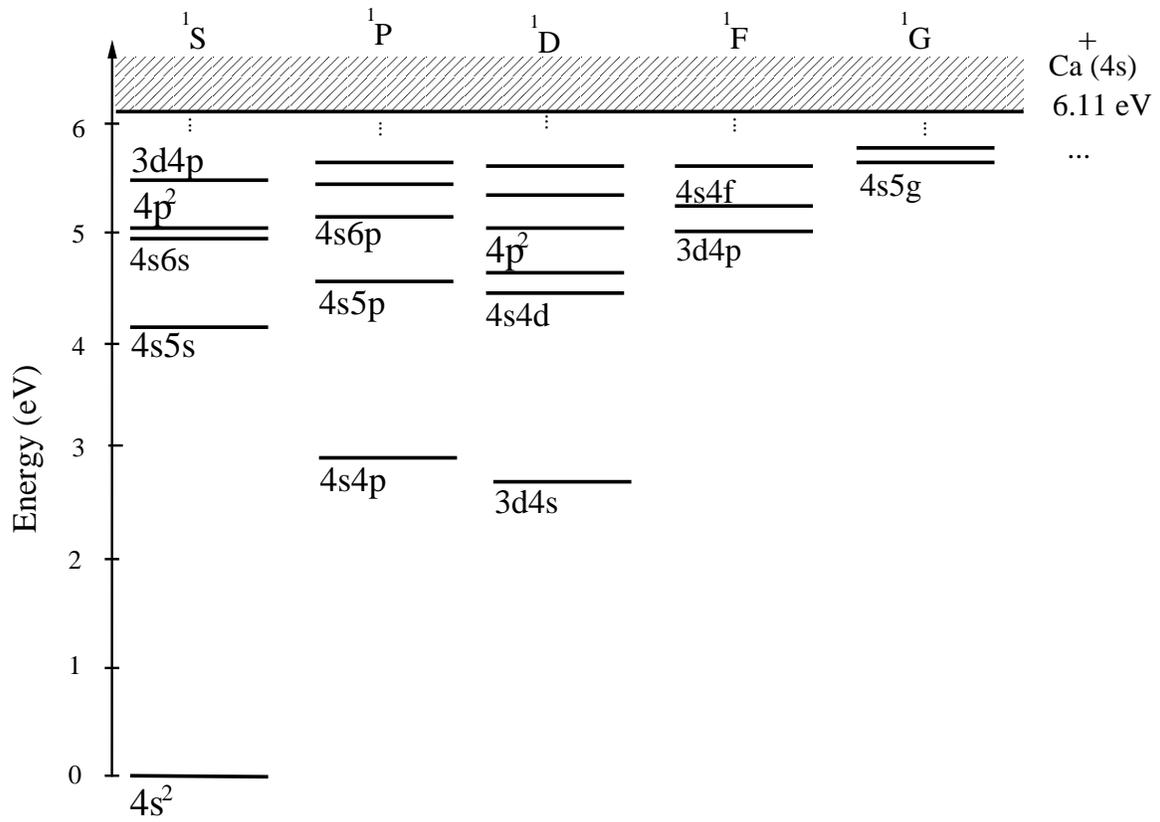}
\caption{Relevant energy levels of Ca. }
\label{fig1}
\end{figure}

\newpage
\begin{figure}
\centering
\includegraphics[width=3.5in,angle=0]{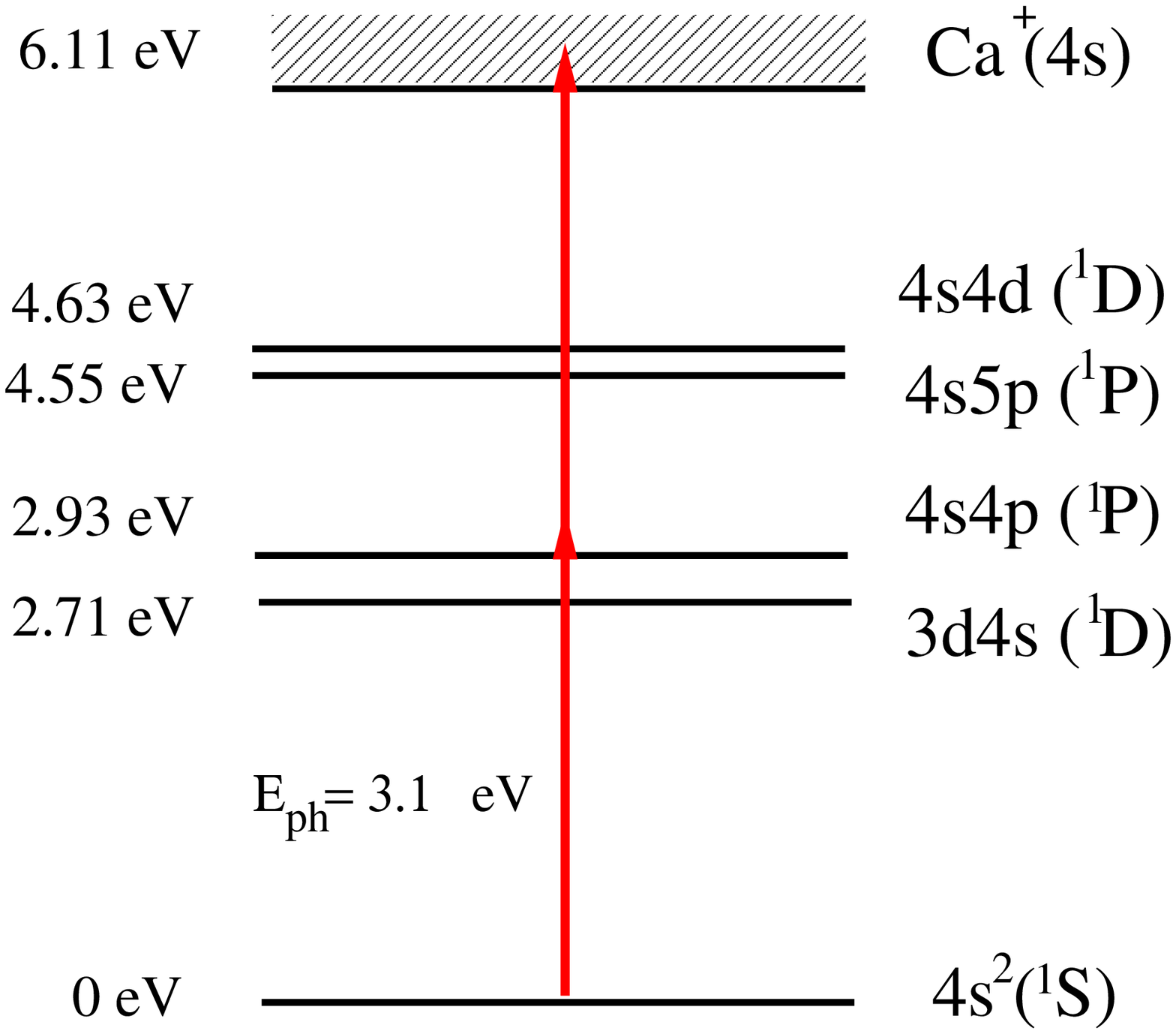}
\caption{
Relevant energy levels for ionization at the photon energy of $ 3.1 $ eV.
The energy detuning from the $4s4p$ $^1P^o$ bound state is $0.17$ eV.}
\label{fig2}
\end{figure}

\newpage
\begin{figure}
\centering
\includegraphics[width=4.5in,angle=0]{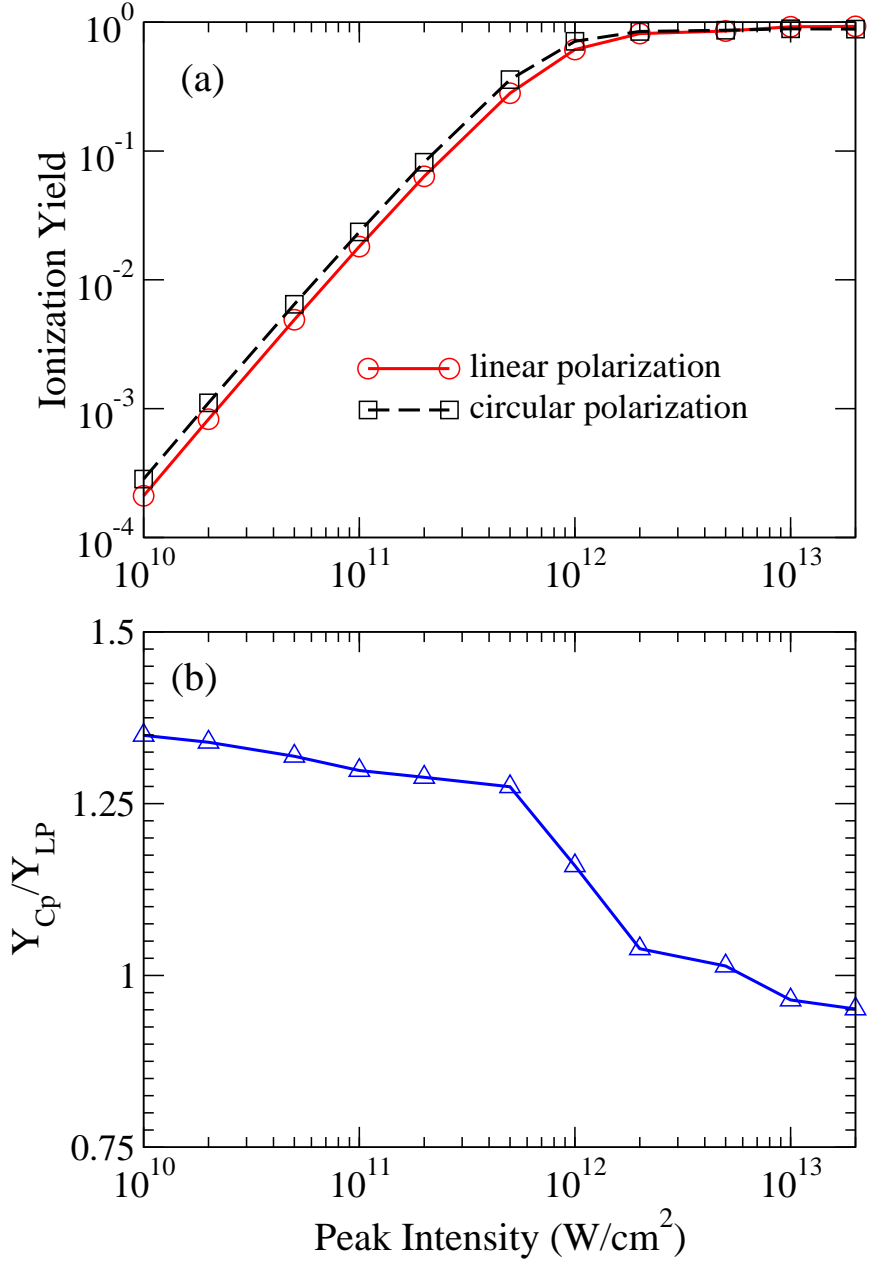}
\caption{(Color Online)
(a) Ionization yield as a  function of peak intensity by the
linearly (solid) and circularly polarized (dashed) laser pulses at the
photon energy of $ 3.1 $ eV.
(b) Ratio of the ionization yield by the circularly polarized laser pulse,
$Y_{CP}$, to that by the linearly polarized laser pulse, $Y_{LP}$.
}
\label{fig3}
\end{figure}

\newpage
\begin{figure}
\centering
\includegraphics[width=6in,angle=0]{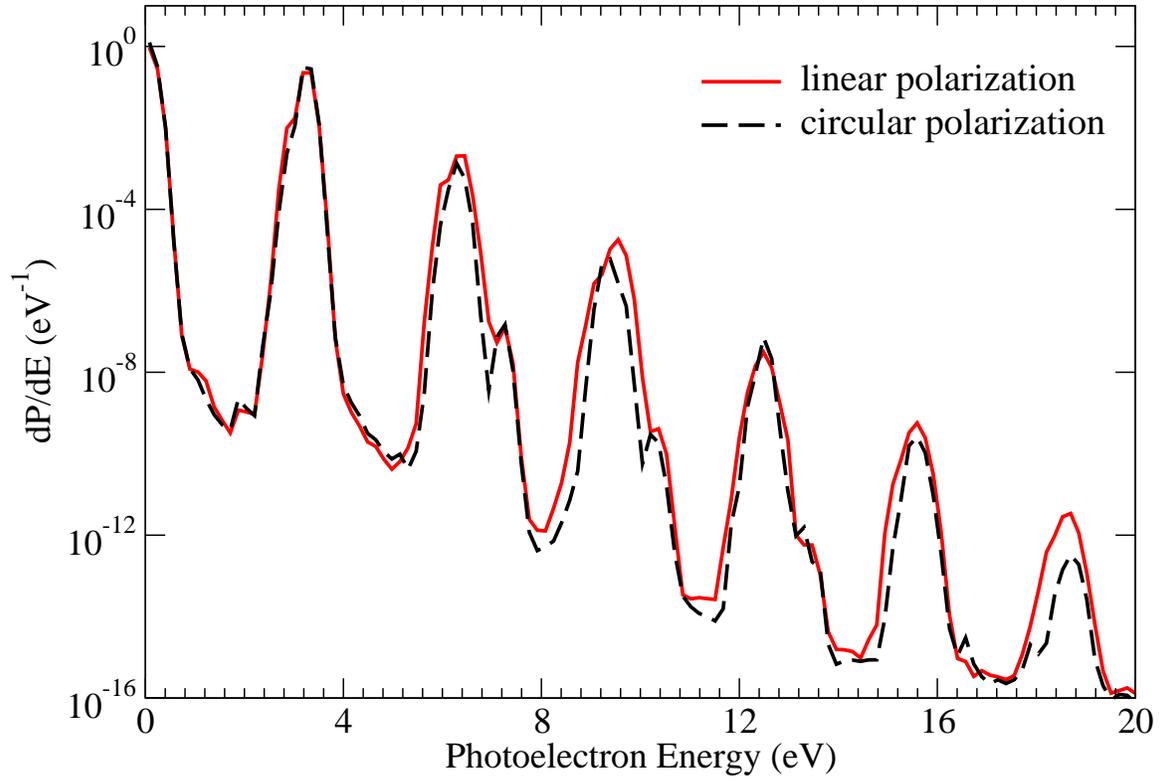}
\caption{(Color Online)
Photoelectron energy spectra by the linearly (solid) and circularly
polarized (dashed) laser pulses at the photon energy  of $ 3.1 $ eV
and the peak intensity of 5$\times$10$^{11}$ W/cm$^2$.
 }
\label{fig4}
\end{figure}

\newpage
\begin{figure}
\centering
\includegraphics[width=6in,angle=0]{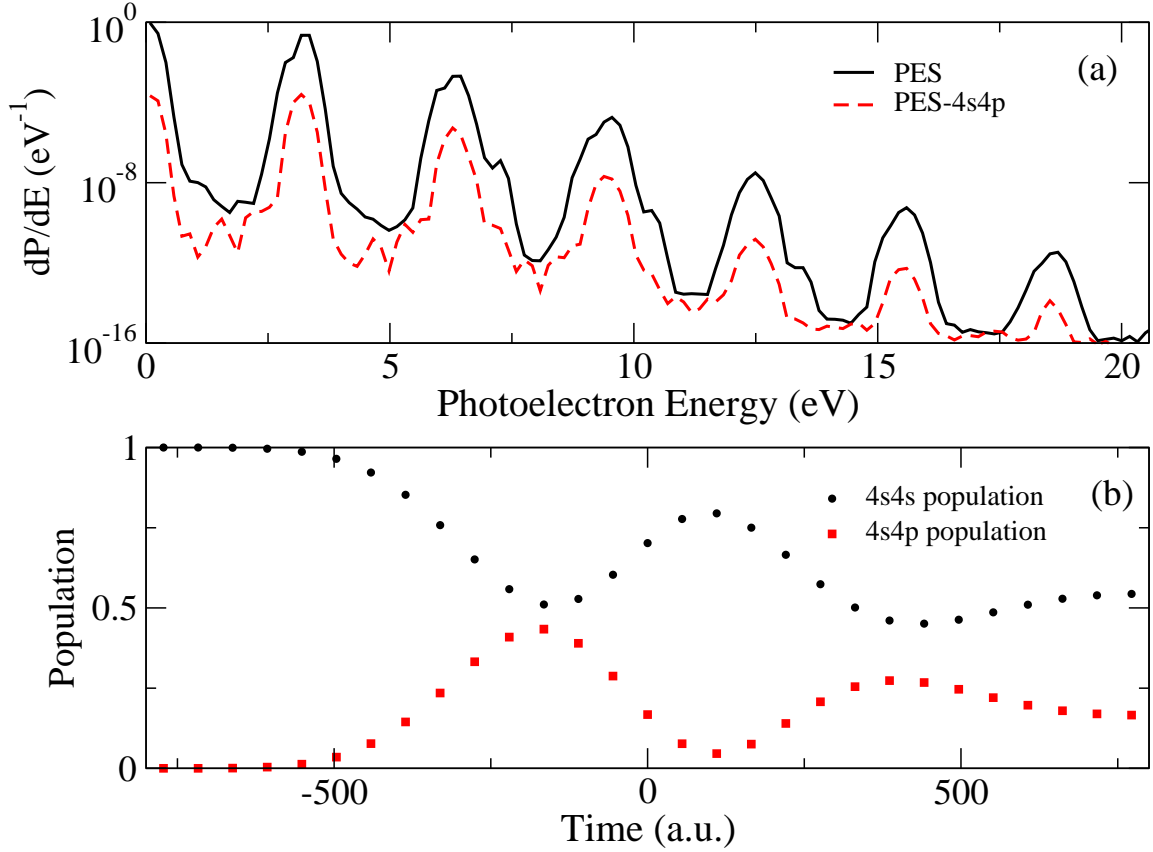}
\caption{(Color Online)
(a) Comparison of the photoelectron energy spectra at the
photon energy of $3.1 $ eV. Solid line represents the result
obtained from the complete atomic basis, while the dashed line represents
the result from the atomic basis without the $4s4p$ $^1P^o$ state
when solving the time-dependent Schr\"odinger equation.
(b) Comparison of the populations of the $4s^2$ $^1S^e$(circles) and
$4s4p$ $^1P^o$ (squares) states as a function of time.
The laser pulse is linearly polarized and the peak intensity is
5$\times$10$^{11}$ W/cm$^2$.}
\label{fig5}
\end{figure}

\newpage
\begin{figure}
\centering
\includegraphics[width=6.in,angle=0]{fig6.eps}
\caption{(Color Online)
Photoelectron angular distributions at the photon energy of 3.1 eV,
corresponding to the (a) first, (b) second, (c) third, and
(d) fourth ATI peaks in Fig. \ref{fig4}.
The laser pulse is linearly polarized and the peak intensity is
5$\times$10$^{11}$ W/cm$^2$.
}
\label{fig6}
\end{figure}

\newpage
\begin{figure}
\centering
\includegraphics[width=3.5in,angle=0]{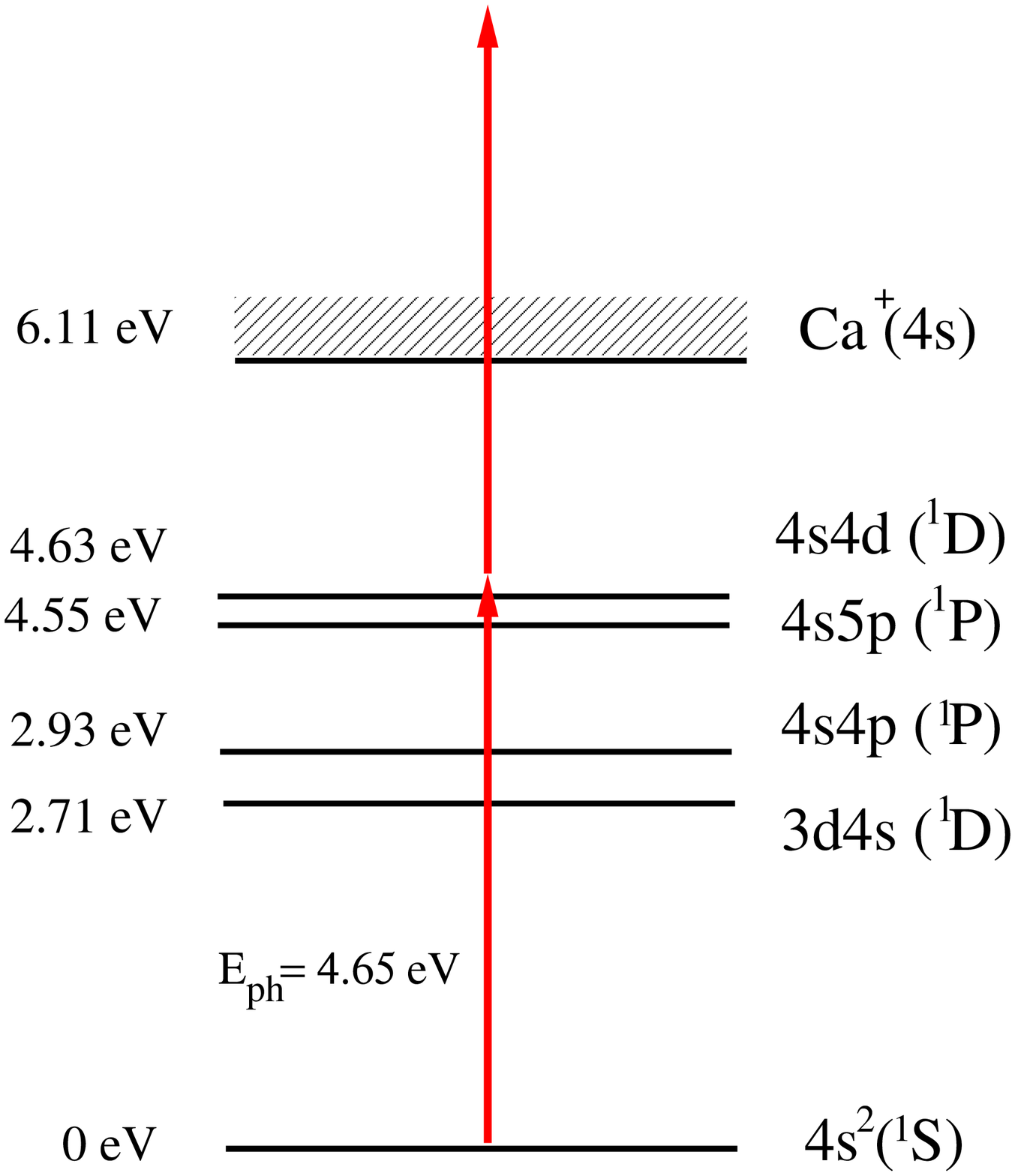}
\caption{
Relevant energy levels for ionization at the photon energy of $ 4.65 $ eV.
The energy detuning from the $4s5p$ $^1P^o$ bound state is $0.1$ eV.
}
\label{fig7}
\end{figure}

\newpage
\begin{figure}
\centering
\includegraphics[width=4.5in,angle=0]{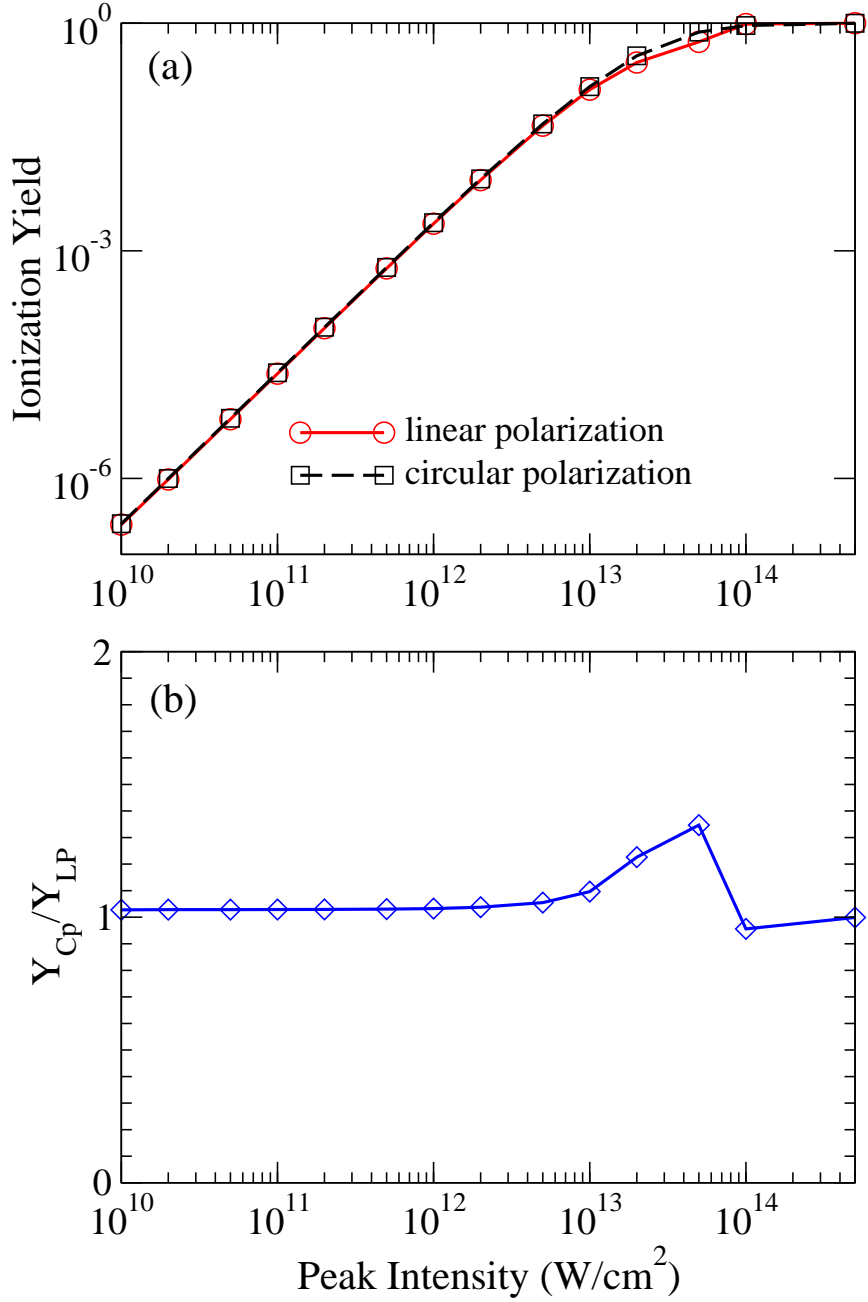}
\caption{(Color Online)
(a) Ionization yield as a  function of peak intensity by the
linearly (solid) and circularly (dashed) polarized laser pulses at the
photon energy  of $ 4.65 $ eV.
(b) Ratio of the ionization yield by the circularly polarized laser pulse,
$Y_{CP}$, to that by the linearly polarized laser pulse, $Y_{LP}$.
}
\label{fig8}
\end{figure}

\newpage
\begin{figure}
\centering
\includegraphics[width=6in,angle=0]{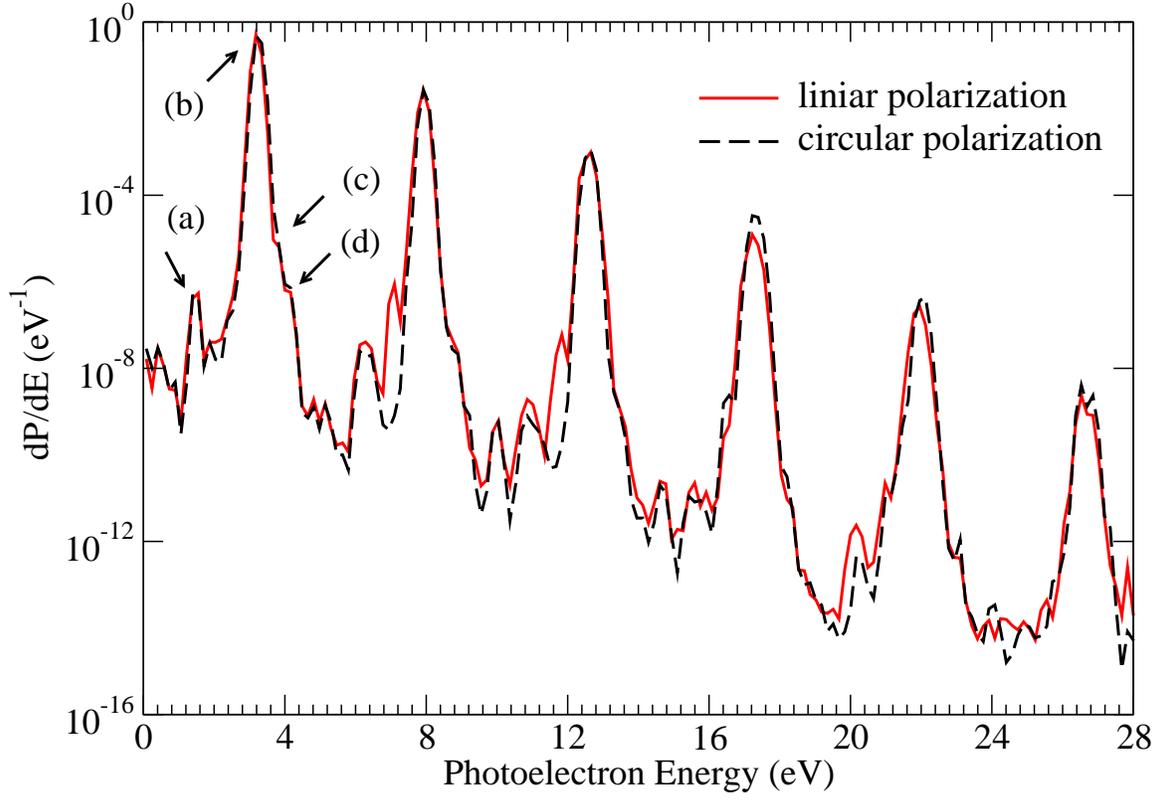}
\caption{(Color Online)
Photoelectron energy spectra by the linearly (solid)  and circularly
polarized (dashed) laser pulses at the photon energy of $ 4.65 $ eV and
the peak intensity of 10$^{13}$ W/cm$^2$.
(a)-(d) represent the substructures.
}
\label{fig9}
\end{figure}

\newpage
\begin{figure}
\centering
\includegraphics[width=5in,angle=0]{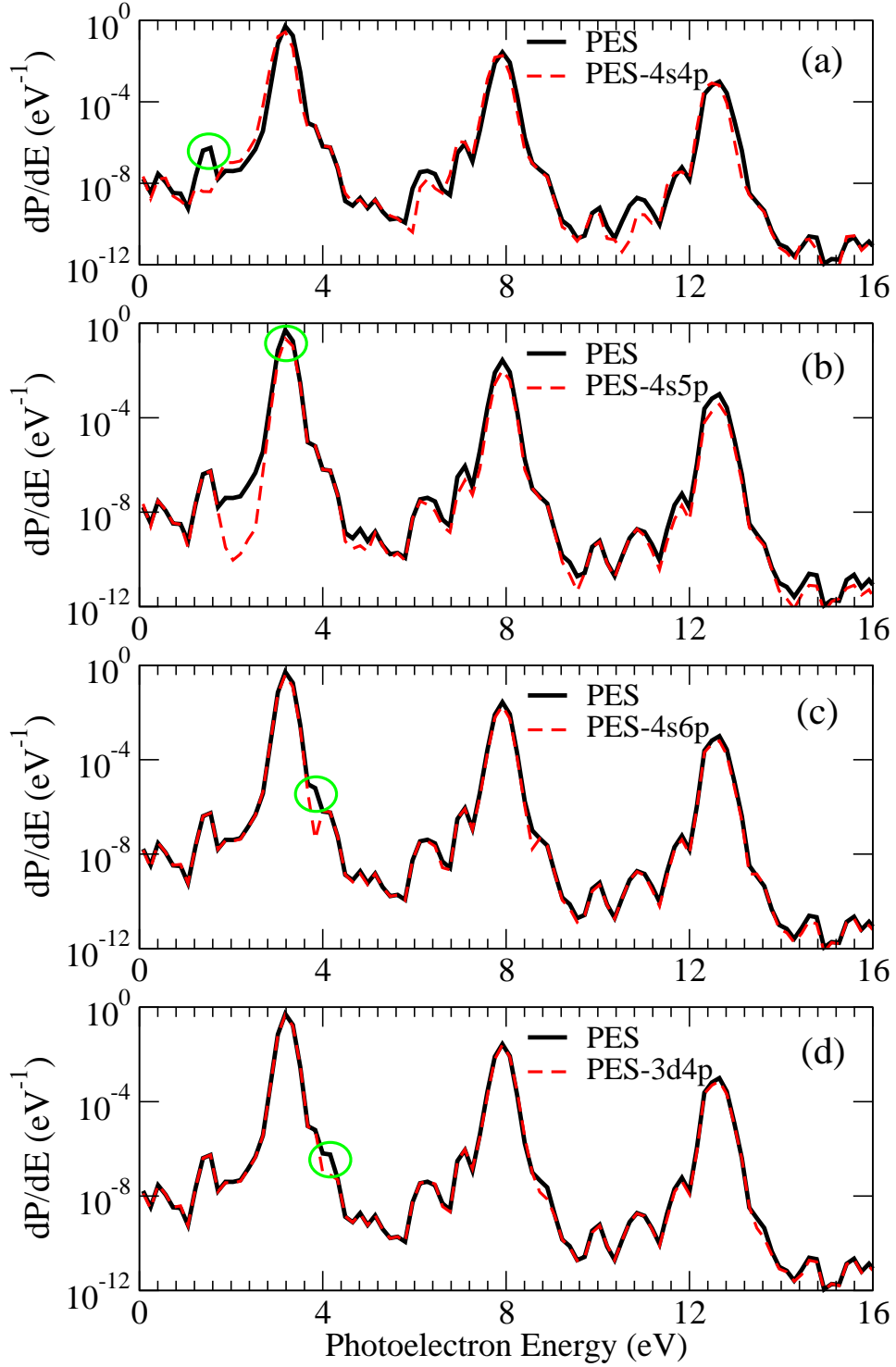}
 \caption{(Color online)
Comparison of the photoelectron spectra at the photon energy of $4.65 $ eV
when the (a) $4s4p$ $^1P^o$, (b) $4s5p$ $^1P^o$, (c) $4s6p$ $^1P^o$ , and
(d) $3d4p$ $^1P^o$ states are artificially removed from the atomic basis
when solving the time-dependent Schr\"{o}dinger equation.
The laser pulse is linearly polarized and the peak intensity is
10$^{13}$ W/cm$^2$.
}
\label{fig10}
\end{figure}

\newpage
\begin{figure}
\centering
\includegraphics[width=6.in,angle=0]{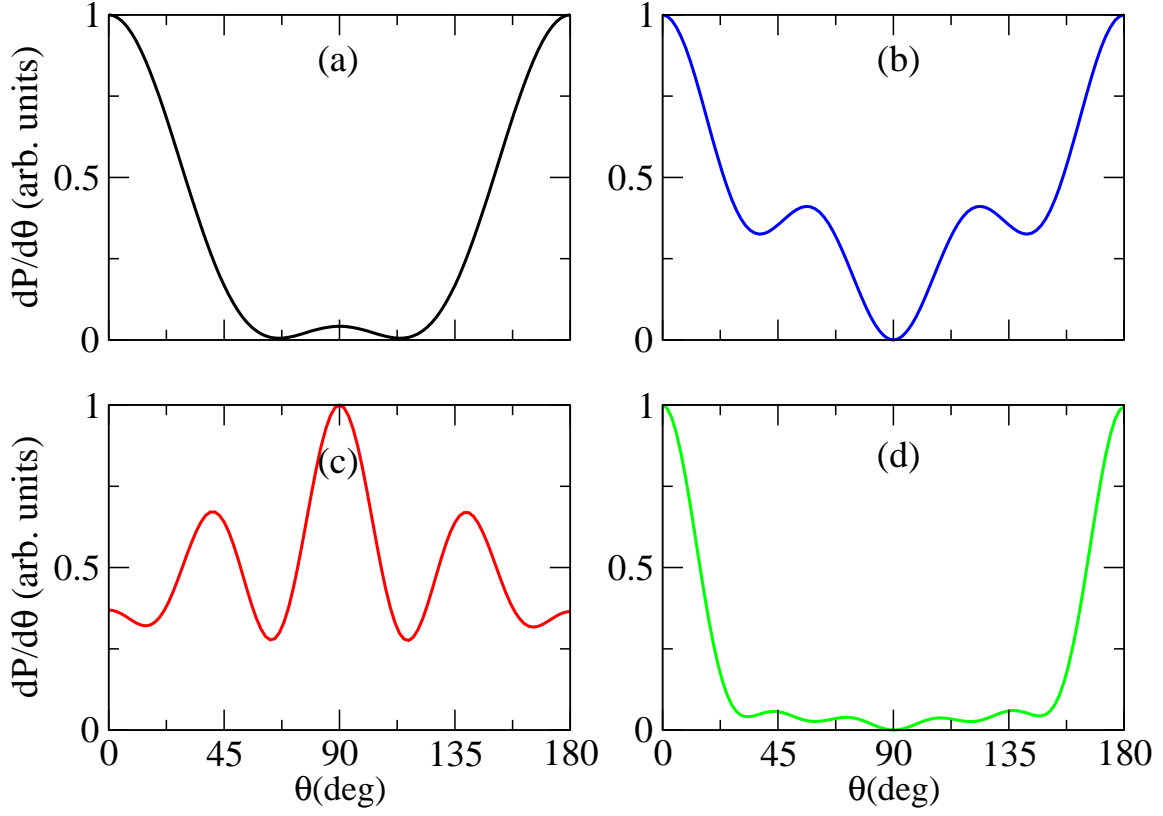}
\caption{(Color Online)
Photoelectron angular distributions at the photon energy of 4.65 eV,
corresponding to the (a) first, (b) second, (c) third, and
(d) fourth ATI peaks in Fig. \ref{fig9}.
The laser pulse is linearly polarized and the peak intensity is
10$^{12}$ W/cm$^2$.
}
\label{fig11}
\end{figure}

\newpage
\begin{figure}
\centering
\includegraphics[width=5in,angle=0]{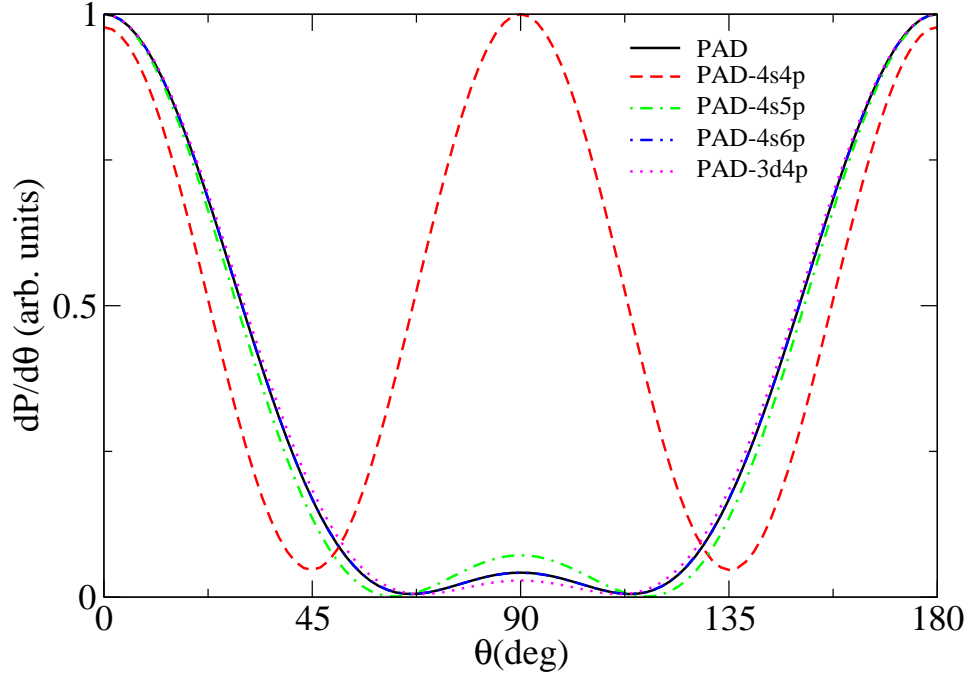}
\caption{(Color Online)
Comparison of the photoelectron angular distributions by a linearly
polarized laser pulse at the photon energy of $4.65 $ eV  when the $4s4p$
$^1P^o$,  $4s5p$ $^1P^o$,  $4s6p$ $^1P^o$, and  $3d4p$ $^1P^o$ bound states of
Ca are artificially removed from the atomic basis when solving the
time-dependent Schr\"{o}dinger equation.
The peak intensity is 10$^{12}$ W/cm$^2$.}
\label{fig12}
\end{figure}

\end{document}